\documentstyle[12pt,titlepage]{article}

\input{tcilatex}

\begin{document}

\begin{center}
{\bf MANYPARTICLE INTERACTIONS AND LOCAL STRUCTURE OF\\[0pt]
THE METALLIC HYDROGEN AT ZERO PRESSURE}

S. D. Kaim$^{1}$, N. P. Kovalenko$^{2}$, E. V. Vasiliu$^{1}$

$^{1}${\it Polytechnic University, 1 Shevchenko av., Odessa, UA-270044,
Ukraine}

$^{2}${\it State University, 2 Dvoryanskaya str., Odessa, UA-270100, Ukraine}
\end{center}

{\small On the basis of the perturbation expansion for electron gas energy
in the third order to the electron-ion potential the pair and irreducible
three-ion interaction potentials in metallic hydrogen are calculated. The
irreducible potential of three-ion interaction has attractive nature at
short interionic separation and oscillates at large ones. The anisotropic
character of the three-ion interaction is shown. The potential relief of the
ions pair relative to the third ion is constructed. This relief has some
potential wells and valleys which connect them. The important role of the
irreducible three-ion interaction in formation of the local order in three-
and four-ion clusters is shown. The potential relief of an equiangular ion
triplet relative to the fourth ion is calculated. This relief has deep
potential well for fourth ion at interionic separation corresponding
interproton separation in molecule H}$_{2}${\small . The quasiclassical
probability of ion transition into this well is evaluated. The life time of
metallic phase of hydrogen relative to the tunneling nucleation of the H}$%
_{2}${\small \ molecules is estimated.}



\begin{center}
{\bf 1. Introduction}
\end{center}

The significance of manyparticle interactions and their role in forming the
structure and properties of various condensed systems are not well known yet 
\cite{1}. For a calculation of manyparticle interaction potentials the
knowledge of the linear and nonlinear response functions of electron gas is
necessary. At present the explicit form of three-pole diagram expression in
ring approximation for homogeneous electron gas is known \cite{2}, \cite{3}.
It was possible to calculate the equilibrium structures and dynamic
properties of simple metal crystals and metallic hydrogen (MH) in the third
order to the electron-ion potential (EIP) \cite{3}-\cite{9}, to study
three-ion interaction potentials in simple metals \cite{3},\cite{7}.

The study of the possibility of metastable state of MH at zero pressure is
of special importance \cite{4}. In a case of solid MH calculations have been
performed by Brovman and Kagan. They calculated the structure, elastic
properties, phonon spectrum of the solid MH at zero pressure and proved the
local stability of the metastable MH phase in framework of the manyparticle
theory of metals \cite{4}. As it turned out in the third order to the EIP
the energy minimum exists for twoparameter hexagonal lattices with
triangular string structure \cite{4}. For all structures with energy minimum
the elementary cell volume was $\Omega_0=20,8a_B^3$ ($a_B$ is the Bohr
radius ) and interproton spacing fixed along Z axis was equal to $d=2,04a_B$%
. The investigations of the higher order approximations did not change
conclusions essentially. It is connected with the fact that eventually the
small parameter in the perturbation expansion for MH energy at P=0
approximately equals to $1/5$ \cite{4},\cite{6}. In such a case the sum of
all perturbation expansion terms with orders $n\geq5$ gives an error as the
dielectric permittivity of the uniform electron gas does \cite{6},\cite{9}.

The estimate of the life time of the metastable MH phase relative to
spontaneous quantum tunneling transition in the insulating phase is among
the most important problems. This estimate cannot be made without
consideration of the specific microscopic mechanism of the nucleation of $%
H_2 $ molecules or $H_2^+$ ions in metallic phase. For the MH as a system in
which the tendency to diatomically ordering exists \cite{8}, the
calculations of interionic interactions potentials and elucidation of this
tendency on the language of groups ion interaction potentials in
configurational space have great interest. In the quantum kinetics of the
new phase formation two characteristic times may be distinguished \cite{10}.
In the first place the time of transition through potential barrier (in our
case the life time of the metallic phase relative to the nucleation of the $%
H_2$ molecule). In the second place the total time of the new phase
formation. As a consequence of a significant difference between electron and
proton masses during the process of molecular hydrogen phase homogeneous
nucleation the characteristic time will be determined by slow ionic motion
in potential relief created by nearest neighbouring ions group. Diatomically
ordering in MH corresponds to the possibility of the drawing together of
ions to the distance corresponding to the interatomic separation in molecule 
$H_2$. Inasmuch the direct interionic interaction corresponds to the Coulomb
repulsion the possibility of drawing ions together may be ensure only by
means of indirect ions interaction through surrounding electron gas.

The purpose of this paper is to consider pair- and three-ion interactions in
metastable MH at P=0. The consideration is based on the manyparticle Brovman
and Kagan approach \cite{3}-\cite{6}. The peculiarities of the potential
relief for ions group are discovered and interpreted from the point of view
manyparticle tunneling formation of the molecular hydrogen phase.

In sec. 2 expressions for many-particle potentials are received. In sec. 3
and sec. 4 results of calculations are shown.

\begin{center}
{\bf 2. Many-body interaction in metals}
\end{center}

In the framework of many-particle theory of non-transition metals the energy
of electron subsystem in the field of fixed ions in the adiabatic
approximation may be written \cite{3}: 
\begin{equation}
E_{e}=\varphi _{0}+\sum_{n}\varphi _{1}({\bf R}_{n})\ +\ \frac{1}{2!}%
\sum_{m\neq {n}}\varphi _{2}({\bf R}_{n},{\bf R}_{m})\ +\ \frac{1}{3!}%
\sum_{m\neq {n}\neq {l}}\varphi _{3}({\bf R}_{n},{\bf R}_{m},{\bf R}_{l})\
+\cdots
\end{equation}
where each term of the (1) series describes interaction of ion groups
through the surrounding electron gas and can be represented as a power
series in the EIP \cite{3}. For example 
\[
\varphi _{2}({\bf R}_{1},{\bf R}_{2})\quad =\quad \sum_{i=2}^{\infty }\Phi
_{2}^{(i)}({\bf R}_{1},{\bf R}_{2}), 
\]
\begin{equation}
\varphi _{3}({\bf R}_{1},{\bf R}_{2},{\bf R}_{3})\quad =\quad
\sum_{i=3}^{\infty }\Phi _{3}^{(i)}({\bf R}_{1},{\bf R}_{2},{\bf R}_{3}),
\end{equation}
etc. The indirect interaction of two ions in second order perturbation
theory in EIP is well known \cite{11} and equal to 
\begin{equation}
\Phi _{2}^{(2)}(R)\quad =\quad \frac{1}{\pi ^{2}}\int\limits_{0}^{\infty
}dqq^{2}\Gamma ^{(2)}(q)|V(q)|^{2}\frac{\sin (qR)}{qR},
\end{equation}
where $V(q)=-\frac{4\pi ^{2}}{q^{2}}$ is the formfactor of the EIP; $\Gamma
^{(2)}(q)$ is the sum of two-pole diagrams.

In the third-order perturbation theory in EIP the indirect pair-ion
interaction is defined by the expression 
\[
\Phi_2^{(3)}(R)\quad=\quad\frac{3}{4\pi^4}\int\limits_{0}^{\infty}
dq_1q_1^2\int\limits_{0}^{\infty}dq_2q_2^2\int\limits_{-1}^{1}dx
V(q_1)V(q_2)V(q_3) \Gamma^{(3)}(q_1,q_2,q_3)\times 
\]
\begin{equation}
\times\frac{\sin(q_1R)}{q_1R},
\end{equation}
where $\Gamma^{(3)}(q_1,q_2,q_3)$ is the sum of three-pole diagrams; $%
q_3=(q_1^2+q_2^2+2q_1q_2x)^{1/2}$.

The potential of indirect three-ion interaction after double integration may
be written \cite{7} 
\[
\Phi _{3}^{(3)}(R_{12},R_{23},R_{13})\quad =\quad \frac{3}{2\pi ^{4}}%
\int\limits_{0}^{\infty }dq_{1}q_{1}^{2}\int\limits_{0}^{\infty
}dq_{2}q_{2}^{2}\times 
\]
\[
\times \int\limits_{-1}^{1}dzV(q_{1})V(q_{2})V(q_{3})\Gamma
^{(3)}(q_{1},q_{2},q_{3})\int\limits_{0}^{1}dx\times 
\]
\[
\times \cos \Biggl(x\biggl(q_{1}R_{12}\frac{R_{12}^{2}+R_{23}^{2}-R_{13}^{2} 
}{2R_{12}R_{23}}+q_{2}R_{23}z\biggr)\Biggr)\times 
\]
\[
\times {J_{0}}\Biggl(q_{1}R_{12}\biggl(1-x^{2}\biggr)^{1/2}\biggl(1-\frac{%
(R_{12}^{2}+R_{23}^{2}-R_{13}^{2})^{2}}{4R_{12}^{2}R_{23}^{2}}\biggr)^{1/2}%
\Biggr)\times 
\]
\begin{equation}
\times {J_{0}\Biggl(q_{2}R_{23}\biggl(1-x^{2}\biggr)^{1/2}\biggl(1-z^{2}%
\biggr)^{1/2}\Biggr)},
\end{equation}
where $J_{0}(x)$ is Bessel function of zeroth order; $z=\cos ({\bf q}_{1},%
{\bf q}_{2})$; $R_{12},\ R_{23},\ R_{13}$- the distances between the
vertices of a triangle formed by the protons.

The calculations of potentials were executed at the Wigner-Zeitz radius $%
r_{S}=1,65$ which corresponds to zero pressure in the zeroth model of a
metal \cite{4}. A permittivity function in the Heldart-Vosko form was
employed. The collections of potential values were calculated on set with
step equal to 1 Bohr radius for each dimension. In the drawings of graphs
was used cubic-spline interpolation.

\begin{center}
{\bf 3. Results}
\end{center}

Pair interionic potential in the third order has a form 
\begin{equation}
\varphi^*(R)=\frac{e^2}{R}+\Phi_2^{(2)}(R)+\Phi_2^{(3)}(R),
\end{equation}
where e is the electron charge.

Figure 1 shows computed potential $\varphi ^{\ast }(R)$ and its components.
The curve 1 corresponds to contribution $\frac{e^{2}}{R}+\Phi _{2}^{(2)}(R)$%
, 2 - $\Phi _{2}^{(3)}(R)$, 3 - $\varphi ^{\ast }(R)$ . Pair interionic
potential $\frac{e^{2}}{R}+\Phi _{2}^{(2)}(R)$ hasn't deep potential well.
Stevenson and Ashcroft \cite{12} have obtained pair interproton potential $%
\frac{e^{2}}{R}+\Phi _{2}^{(2)}$ for MH at $r_{S}=1,6$ analogous to showed
in Fig. 1 (curve 1). Indirect interionic interaction $\Phi _{2}^{(3)}(R)$
has attractive nature and forms potential well for nearest neighbouring ion
and minimum in the repulsive part of the potential $\varphi ^{\ast }(R)$.
The decrease of the density leads to strong increase of the deep of minimum
in the repulsive part of the potential $\varphi ^{\ast }(R)$. The curve 1 at
the Fig. 2 corresponds to potential $\varphi ^{\ast }(R)$ at $r_{S}=1,72$.
At $r_{S}<1,65$ the minimum in the repulsive part of $\varphi ^{\ast }(R)$
turns shallow and its position shifts towards smaller R and at $r_{S}=1,55$
this minimum disappears (curve 2 in Fig. 2). The position of minimum in the
repulsive part of $\varphi ^{\ast }(R)$ in the Fig. 1 corresponds to
separation $R=1,6a_{B}$. It should be mentioned that the separation between
nuclei in $H_{2}$ molecule is equal to $1,4a_{B}$ and binding length in $%
H_{2}^{+}$ ion is equal to $2a_{B}$ \cite{13}. The formation of the minimum
in the dependence at a distance which is smaller than the interproton
average distance and strong dependence of the deep of this minimum from
density of MH may be treated as the tendency to diatomic ordering in
electron-proton plasma.

The results of calculation of the irreducible three-ion interaction
potential $\Phi _{3}^{(3)}$ for $r_{S}$ = 1,65 are presented in Fig. 3 in
form of the potential relief. The minimal interproton separation in metallic
phase at P=0 equals to $2,04a_{B}$ \cite{4}. Two ions are placed in ordinate
axis on distance $2a_{B}$ and third ion is sited in plane XOY. For small
distances the potential $\Phi _{3}^{(3)}(2,R_{23},R_{13})$ has attractive
character. The potential $\Phi _{3}^{(3)}(2,R_{23},R_{13})$ has anisotropic
form and its most rapidly changing takes place in OX direction. For large
distances the potential $\Phi _{3}^{(3)}$ has oscillatory character. Note
that $\Phi _{3}^{(3)}(0,0,0)\simeq -1\ Ry$.

Fig. 4 (geometry coincides with Fig. 3) shows potential relief for the 3-rd
ion in the fields of two other ions with pair interactions taken into
account, that is the potential $\varphi ^{\ast }(R_{23})+\varphi ^{\ast
}(R_{13})$. The minima B and D and a valley which connects them correspond
to the possible positions of the nearest neighbouring ion in the field of
two fixed ions (see Fig. 4). It should be noted that ion transition from
minimum B to minimum D along the valley entails the crossing the potential
barrier with height $\sim 1200\ K$. This situation remains unchanged when
irreducible three-proton interaction is taken into account because of the
short-acting character of potential $\Phi _{3}^{(3)}$ . The quantum
consideration of ion motion in the field of two fixed ions results to
conclusion about splitting of energy level for ion which is the nearest
neighbour for pair considered. The local minima A and C in Fig. 4 correspond
to distances to ion 1 and 2 which are less than interionic average distance.

Fig. 5 shows the potential relief which is created ions pair and taking into
account pair- and three-ion interactions. It is described by function $%
\varphi ^{\ast }(R_{23})+\varphi ^{\ast }(R_{13})+\Phi
_{3}^{(3)}(2,R_{23},R_{13})$. The change of local minima A,B,C,D positions
as compared with Fig. 4 is unimportant. As a consequence of short-range and
anisotropic character of potential $\Phi _{3}^{(3)}$ the depth of minimum A
changes markedly and the depth of minimum B increases a bit. The ion wave
function in potential relief which is shown in Fig. 5 has nonzero value near
local minima A and C. This means that ions have finite probability draw
together to separation of the order of internuclei distances in $H_{2}$ or $%
H_{2}^{+}$. From comparison of Fig. 1 and Fig. 5 the conclusion follows that
the presence of third ion essentially increases this probability.

Fig. 6 shows computed distribution of conditional probability density for
ion positions in ions pair field with pair and three-ion interactions taken
into account in Boltzmann approximation at temperature T = 1000 K. In this
approximation distribution of the conditional probability density may be
described in the form 
\begin{equation}
F_{1}({\bf R}_{1}|{\bf R}_{2},{\bf R}_{3})\sim \exp \Biggl(-(\varphi ^{\ast
}(R_{12})+\varphi ^{\ast }(R_{13})+ \\
\Phi _{3}^{(3)}(R_{12},2,R_{13}))/kT\Biggr),
\end{equation}
where $k$ is Boltzmann constant. Local minima of the potential relief at
Fig. 5 correspond to peaks of the function $F_{1}({\bf R}_{1}|{\bf R}_{2},%
{\bf R}_{3})$.

It is interesting to consider the potential relief which is created by group
of ions. Fig. 7 shows potential relief which three ions forming equiangular
triangle with side $2a_{B}$ and placing in XOY plane create in the same
plane towards fourth ion. In Fig. 7 the pair-ion interaction is taken into
account and potential relief assigns to function $\varphi ^{\ast
}(R_{14})+\varphi ^{\ast }(R_{24})+\varphi ^{\ast }(R_{34})$. It is easy to
see two minima A and B and saddle point C in XOY plane.

Potential relief which is shown in Fig. 8 (the geometry is the same as in
Fig. 7 ) and which is described by function 
\[
\varphi ^{\ast }(R_{14})+\varphi ^{\ast }(R_{24})+\varphi ^{\ast
}(R_{34})+\Phi _{3}^{(3)}(2,R_{24},R_{14})+ 
\]
\begin{equation}
\Phi _{3}^{(3)}(R_{24},2,R_{34})+\Phi _{3}^{(3)}(R_{14},R_{34},2)
\end{equation}
takes into account three-ion interactions in the cluster of four ions.
Three-ion interactions transform the saddle C into local minimum and minima
A and B become deeper considerably. This means that an increase of the
particles number in cluster lowers the potential barrier to diatomic
ordering.

If three ions are situated so as is shown in Fig. 8 and the fourth ion is
placed on Z-axis, the dependence of potential with pair- and three-ions
interactions taken into account is shown in Fig. 9. The presence of deep
minimum at $z=a_{B}$ is characteristic peculiarity of this dependency.
Second minimum at $z=3,5a_{B}$ corresponds to equilibrium position of fourth
ion.

The quantum description of ion motion in two-well potential of Fig. 9 yields
the conclusion about a possibility of tunneling transition from well A to
well B. The probability of ion transition per 1 s is equal to product of the
ion oscillation frequency in well A and transmission coefficient through
potential barrier. The transmission coefficient can be estimated in
quasiclassical approach. The ion frequency of zero-oscillation in the well A
is equal to $\nu \ =\ 3,61\cdot 10^{13}$ Hz. The transmission coefficient
through potential barrier $D\ =\ 3,85\cdot 10^{-12}$. Then ion life time $%
\tau $ in well A is equal to $\tau \ =\ (D\nu )^{-1}\ =\ 0,0072$ s. If ion
transition into well B takes place the conditions for electrons localization
on ion pairs will be created. Taking into consideration also small
characteristic time of an electron subsystem it is possible to consider
obtained time as characteristic time of the tunneling nucleation of $H_{2}$
molecule in MH at zero pressure. If the identity ions is taken into account
it is possible to make a conclusion about manyionic tunneling mechanism for
metallic hydrogen transition into molecular phase.

The quantum Monte-Carlo simulations of the molecular hydrogen \cite{14} and
calculations of the diatomic phase energy as function of the interproton
separation \cite{8} indicate that at chosen density the molecular unit
interproton separation is somewhat less than the neutral molecule separation 
$1,4a_{B}$. We calculated the potential relief which is created by ions pair
with interproton separation $1,5a_{B}$ (Fig. 10,11). Fig. 10 shows potential
relief for the third ion in the field of two other ions with pair-ion
interaction in third order perturbation theory (analogous to Fig. 4). Fig.
11 (geometry coincides with Fig. 10) shows potential relief for the third
ion with pair and irreducible three-ion interactions taken into account,
that is the potential $\varphi ^{\ast }(R_{12})+\varphi ^{\ast
}(R_{2}3)+\Phi _{3}^{(3)}(1.5,R_{13},R_{23})$. As is easily seen from Fig.
10,11 , the irreducible three-ions interaction essentially deepens the
potential well A ( Fig. 11 is analogous to Fig. 5 ). The comparison of Fig.
5 and Fig. 11 yields the conclusion about strong dependence on the
interproton separation $R_{12}$ of the probability of triplet ions drawing
together to the interproton separation as in molecule $H_{2}$ . The time of
the proton transition from well B into well A (Fig. 11) cannot be considered
as characteristic time of the tunneling nucleation $H_{2}$ molecule in
metallic phase since interproton separation $R_{12}=1,5a_{B}$ is less than
minimal interproton separation in MH phase at P=0 which equals to $2,04a_{B}$
\cite{4}.

\begin{center}
{\bf 4. Discussion and conclusions}
\end{center}

Our consideration is based on the assumption that metastable MH phase exists
at P=0. The calculations show that irreducible three-ion indirect
interactions have decisive influence on the local structure of the ionic
subsystem. The calculations of the possible MH structures at P=0 show the
tendency for MH to crystallize in triangular string with two-dimensional
periodicity structure \cite{4}. Our calculations of the potential relief
which three ions create towards fourth ion confirm this conclusion and show
that taking into account $\Phi _{3}^{(3)}$ really leads to formation of the
hexagonal structure in the plane of four ions arrangement. Taking into
account $\Phi _{3}^{(3)}$ leads to transformation of the saddle C (Fig. 7)
in local minimum (Fig. 8). Minimum C is formed along direction which forms
angle $60^{0}$ with OX axis (the curves on Fig. 7, 8 are drawn with a step $%
15^{0}$ ).

It is possible to observe an interesting peculiarity of the potential relief
which ion pair creates towards third ion (Fig. 4,5,6): local minima B and D
and valley which connects them and has saddle point. To make transition from
minimum B to D it is necessary to cross over the potential barrier. This
means that energy levels of the third ion in field of two fixed ions are
split. The quantum effect of energy levels splitting will be exercised at
temperatures which are less than potential barrier height. It should be
noted here that as a consequence of short- range character of potential $%
\Phi _{3}^{(3)}$ it has weak influence on this effect (Fig. 4,5). It can be
supposed that the existence of two-level subsystems in amorphous solids at
low temperature is somehow the display of collective quantum oscillatory
states of three particle groups.

In this study the peculiarities of the pair- and three-ion interactions are
interpreted as the vestiges of the molecular phase in metastable MH. These
peculiarities occur in metastable metallic phase and are absent at megabar
pressure (Fig. 2). This means that electron-proton system possesses a
self-organization effect of the ''preparing'' a transition to the molecular
phase. Such effect increases sharply with decreasing of the electron gas
density.

The probability of ions drawing together to the interproton separation as in
molecule $H_2$ depends essentially on the electron gas density. With a
density decreasing the probability of the manyionic tunneling nucleation of
the molecules $H_2$ sharply increases. Therefore some interest has a study
of the life time of metastable MH relative to the homogeneous nucleation of
the molecular phase in pressure interval $0<P<P_t$ ($P_t$ is the pressure of
the molecular hydrogen transition in metallic phase).

The calculation of the life time of the MH as macroscopic system needs the
consideration both the mechanism of homogeneous nucleation of molecular
phase and kinetics of formation of the insulating phase droplets in the
metastable metal. In this paper we did not consider a kinetic stage. Such
consideration must include a search of volume and surface free energy
minimum of heterogeneous system consisting of metastable metal and
insulating droplets. In accordance with the classical and quantum theories
of the new phase formation \cite{10},\cite{15},\cite{16} calculated in this
paper life time of metallic phase relatively tunneling nucleation of the $%
H_{2}$ molecule may differ from life time of MH as macroscopic system by a
factor $\sim $ 3-5. The new experimental data on structures and phase
diagrams for molecular hydrogen and deuterium at megabar pressure \cite{17}, 
\cite{18} may be very useful by consideration of the kinetic stage of the
formation and growth of the insulating phase droplets in the metastable MH.

\begin{center}
{\bf Acknowledgements}
\end{center}

The authors are grateful to Prof. Yu. P. Krasny, Prof. Z. A. Gurskii and
Prof. M. V. Vavrukh for stimulating discussion. This work was supported, in
part, by International Soros Education Program of the International
Renaissance Foundation through grants No APU061042 and No SPU062029.

\begin{center}
{\bf Figure captions}
\end{center}

Fig. 1. Potential $\varphi ^{\ast }(R)$ and its components for $r_{S}=1,65$.

Fig. 2. Potential $\varphi ^{\ast }(R)$ for $r_{S}=1,72$ and $r_{S}=1,55$.

Fig. 3. Irreducible three-ion interaction potential $\Phi
_{3}^{(3)}(2,R_{13},R_{23})$.

Fig. 4. Potential relief for the 3-rd ion in the field of two ions with pair
interactions taken into account.

Fig. 5. Potential relief for the 3-rd ion in the field of two ions with
pair- and three-ion interactions taken into account.

Fig. 6. Distribution of conditional probability density for ion positions in
ions pair field. \newline
Fig. 7. Potential relief which three ions create toward fourth ion (pair
interactions taken into account).

Fig. 8. Potential relief which three ions create toward fourth ion (pair-
and three-ion interactions taken into account).

Fig. 9. The dependence of potential $\Phi (z)$.

Fig. 10. Potential relief for the 3-rd ion in the field of two ions
(interionic separation is equal to $1,5a_{B}$) with pair interactions taken
into account.

Fig. 11. Potential relief for the 3-rd ion in the field of two ions
(interionic separation is equal to $1,5a_{B}$) with pair- and three-ion
interactions taken into account.

\end{document}